\documentclass[%
reprint,
 amsmath,amssymb,
prb,
]{revtex4-1}
\usepackage{mathrsfs}
\usepackage{graphicx}% Include figure files
\usepackage{dcolumn}% Align table columns on decimal point
\usepackage{bm}% bold math
\usepackage[breaklinks=true,colorlinks=true,linkcolor=blue,urlcolor=blue,citecolor=blue]{hyperref}
\usepackage{color}
\newcommand{\ud}{\ensuremath{\mbox{d}}}

\begin{document}
\preprint{APS/123-QED}

\title{A discussion of $Bl$ conservation on a two dimensional magnetic field plane in watt balances}% Force line breaks with \\
\author{Shisong Li}
 \altaffiliation[]{Department of Electrical Engineering, Tsinghua University, Beijing 100084, China}%Lines break automatically or can be forced with \\
 \email{leeshisong@sina.com}
 \author{Wei Zhao}
 \author{Songling Huang}
% \author{Jiansheng Yuan}

\date{\today}% It is always \today, today,

\begin{abstract}
The watt balance is an experiment being pursued in national metrology institutes for precision determination of the Planck constant $h$. In watt balances, the $1/r$ magnetic field, expected to generate a geometrical factor $Bl$ independent to any coil horizontal displacement, can be created by a strict two dimensional, symmetric (horizontal $r$ and vertical $z$) construction of the magnet system. In this paper, we present an analytical understanding of magnetic field distribution when the $r$ symmetry of the magnet is broken and the establishment of the $Bl$ conservation is shown. By using either Gauss's law on magnetism with monopoles or conformal transformations, we extend the $Bl$ conservation to arbitrary two dimensional magnetic planes where the vertical magnetic field component equals zero. The generalized $Bl$ conservation allows a relaxed physical alignment criteria for watt balance magnet systems.
\end{abstract}

\pacs{Valid PACS appear here}% PACS, the Physics and Astronomy
                             % Classification Scheme.
%\keywords{Suggested keywords}%Use showkeys class option if keyword
                              %display desired
'\maketitle

%\tableofcontents
\section{Introduction}
A handful of national metrology institutes (NMIs) have designed and constructed watt balances \cite{kibble, NPL12, NIST14, NRC14, METAS13, BIPM13, LNE14}, a high-precision force comparator leading the effort in redefining the SI unit of mass, the kilogram (kg), in terms of the Planck constant $h$, expected to culminate in 2018 \cite{CIPM}. A watt balance experiment is comprised of two modes of operation, i.e., weighing mode and velocity mode. In the weighing mode, the downwards gravitational force exerted on a test mass is compensated by an upward electromagnetic force generated by a current-carrying coil in a magnetic field. The electromagnet's geometrical factor $(Bl)_w$ is measured as
\begin{equation}
(Bl)_w=\frac{mg}{I},
\end{equation}
where $m$ denotes the test mass, $g$ the local gravitational acceleration, $B$ the magnetic flux density at the coil position, $l$ the coil wire length, and $I$ the current through the coil.
In the velocity mode, the coil is moved in the magnetic field at a constant, vertical velocity $v$, yielding an induced voltage $U$, and the geometrical factor $(Bl)_v$ is calibrated as
\begin{equation}
(Bl)_v=\frac{U}{v}.
\end{equation}
By comparing the measured values of $I$ and $U$ to electrical quantum standards, i.e., the Josephson voltage standard\cite{JVS} and the quantum hall resistance standard\cite{QHR}, the Planck constant $h$ is determined by a ratio of mechanical power and electrical power, i.e.,
\begin{equation}
h=h_{90}\frac{(Bl)_w}{(Bl)_v}=h_{90}\frac{mgv}{UI},
\end{equation}
where $h_{90}\equiv6.626068854...\times10^{-34}$Js is a derived value of the Planck constant by the 1990 conventional electrical units \cite{1990}.
In order for each watt balance to qualify for the redefinition of the kilogram in 2018, each instrument must accurately measure $h$ with a relative uncertainty of several parts in $10^{8}$.

In a good measurement design, the measured quantity should be insensitive to as many environmental variables as possible. One such approach in watt balance theory is to design a $1/r$ ($r$ is the coil radius) magnetic field at the weighing position by either a electromagnetic system \cite{NIST14} or a symmetrical permanent magnet \cite{Li13}. It has been mathematically proven that $Bl$ is conserved in a pure $1/r$ magnetic field, which is therefore insensitive to any effects caused by undesired horizontal coil displacements or radially geometric fluctuations of the coil due to thermal expansion \cite{Li15}.
However, the generation of an ideal $1/r$ magnetic field strongly depends on concentrically aligning the inner and outer yoke, i.e., achieving horizontal $r$ symmetry, proven to be a formidable challenge in practice. Our curiosity is whether $Bl$ is still conserved when the $r$ symmetry of the magnetic field is broken.
To answer this question, we first give an analytical analysis of the asymmetrical magnetic field on a two dimensional field plane when the inner yoke and outer yoke are not concentric. The $Bl$ is proven to remain a constant as in the symmetrical example.
Without losing generality, the $Bl$ conservation can be further extended by Gauss's law of magnetism on any two dimensional field planes by employing the magnetic monopole model \cite{Nature}. This conclusion can be also explained by conformal transformations. The $Bl$ conservation concept relaxes the design and alignment specifications required for a watt balance magnet.

The outline of the rest of this article is organized as follows: In section \ref{sec2}, a typical watt balance magnet and the ideal symmetrical $1/r$ case is reviewed. In section \ref{sec3}, an analysis of the asymmetrical case when the inner yoke and outer yoke are not concentric is presented, and the $Bl$ conservation is shown. A more generalized extension of $Bl$ conservation by employing the Gauss's law of magnetism and the conformal transformation is shown in section \ref{sec4}.

\section{Review of an ideal case with $1/r$ magnetic field}
\label{sec2}
\begin{figure}[tp!]
 \centering
 \includegraphics[width=0.22\textwidth]{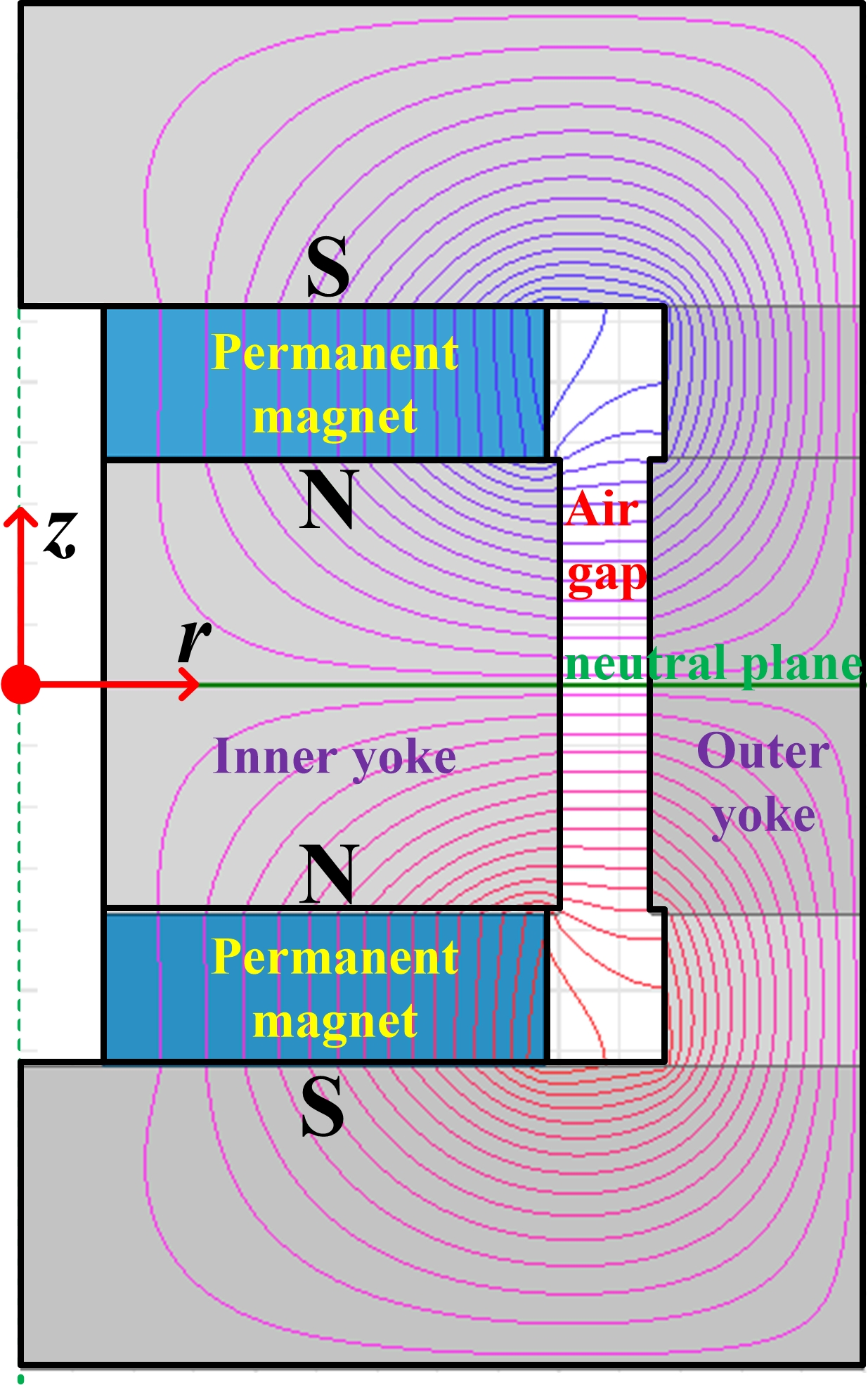}
 \caption{A typical construction of the watt balance magnet (half cross sectional view). N and S denote the north and south poles of the permanent magnet. The magnetic flux lines through the circuit are shown. A similar watt balance magnet is employed in many NMI watt balances, e.g., the BIPM watt balance \cite{BIPM13}, METAS-II watt balance \cite{METAS13}, NIST-4 watt balance \cite{NISTmag}, KRISS watt balance \cite{Kriss}.}
 \label{fig1}
\end{figure}

The analysis in this article is based on one of the most popular watt balance magnets employed \cite{BIPM13, METAS13, NISTmag, Kriss}. Figure \ref{fig1} shows the construction of the magnet with a symmetrical structure in both the $r$ and $z$ axes. The magnetic flux, generated by two permanent magnet disks oriented in repulsion, is guided by yokes of high permeability material through an air gap radially separating the inner and outer yokes.

Based on magnetostatic equations, $\nabla\cdot\bf B=0$ and $\nabla\times\bf B=0$, the magnetic field in the air gap can be described in the cylindrical
coordinate as \cite{LiYuan}
\begin{equation}
\frac{1}{r}\frac{\partial [rB_r(r,\phi, z)]}{\partial r}+\frac{1}{r}\frac{\partial B_\phi(r,\phi, z)}{\partial \phi}+\frac{\partial B_z(r,\phi, z)}{\partial z}=0,
\label{eq.maxI}
\end{equation}

\begin{equation}
\frac{1}{r}\frac{\partial B_z(r,\phi,z)}{\partial \phi}-\frac{\partial B_\phi(r,\phi,z)}{\partial z}=0,
\label{eq.MaxII1}
\end{equation}

\begin{equation}
\frac{\partial B_r(r,\phi,z)}{\partial z}-\frac{\partial B_z(r,\phi,z)}{\partial r}=0,
\label{eq.MaxII2}
\end{equation}

\begin{equation}
\frac{1}{r}\frac{\partial [rB_\phi(r,\phi,z)]}{\partial r}-\frac{1}{r}\frac{\partial B_r(r,\phi,z)}{\partial \phi}=0.
\label{eq.MaxII3}
\end{equation}

For the ideal case, the magnet shown in figure \ref{fig1} is symmetrically constructed in both $r$ and $z$ directions, thus we have $B_\phi=0$, $\partial B_r(r,\phi, z)/\partial\phi=0$ and $\partial B_z(r,\phi, z)/\partial\phi=0$.
Then equations (\ref{eq.maxI})-(\ref{eq.MaxII3}) turns to
\begin{equation}
\frac{1}{r}\frac{\partial [rB_r(r,z)]}{\partial r}+\frac{\partial B_z(r,z)}{\partial z}=0,
\label{eq.MaxI2D}
\end{equation}
\begin{equation}
\frac{\partial B_r(r,z)}{\partial z}-\frac{\partial B_z(r,z)}{\partial r}=0.
\label{eq.MaxII2D}
\end{equation}

Equations (\ref{eq.MaxI2D}) and (\ref{eq.MaxII2D}) show that the two magnetic flux density components, $B_r$ and $B_z$, are coupled. In watt balance operations, the weighing is conducted at the neutral plane of the magnet, i.e. $z=0$. Since the vertical magnetic field component is zero on the $z=0$ plane, the magnetic flux density contains a pure $B_r$ component and is solved as
\begin{equation}
B_r(r,0)=\frac{r_cB_r(r_c,0)}{r}=\frac{\mathscr{B}}{r},
\end{equation}
where $r_c$ is the coil radius and $\mathscr{B}$ is the product of $B_r(r,0)$ and $r$. It has been mathematically proven in \cite{Li15} that the $Bl$ at the neutral plane is independent from the coil horizontal position and the coil radius, i.e.,
\begin{equation}
(Bl)_{z=0}=2\pi r_cB_r(r_c,0)=2\pi\mathscr{B}.
\end{equation}

\section{$Bl$ conservation in asymmetrical magnetic field}
\label{sec3}
\subsection{Analytical expression of the magnetic field}
\begin{figure}[tp!]
 \centering
 \includegraphics[width=0.45\textwidth]{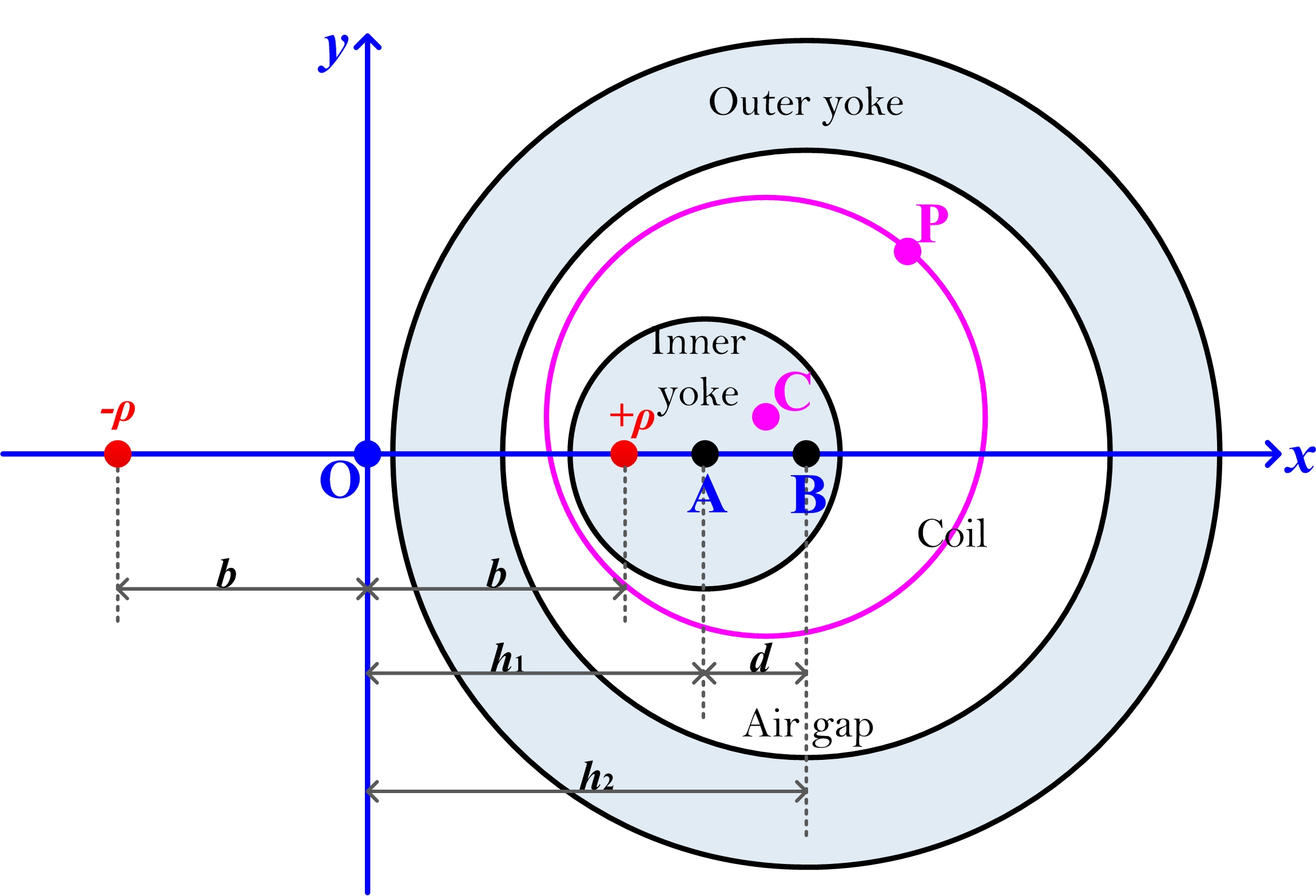}
 \caption{The magnet and coil dimensions on the $z=0$ plane. A and B are centers of the inner yoke and outer yoke respectively while C is the coil center. }
   \label{fig2}
\end{figure}

Here we give an analytical solution of the magnetic field when the inner and outer yokes are not concentrically aligned. The magnet and coil geometric relationships on the $z=0$ plane have been shown in figure \ref{fig2}. Two boundary lines, the outer radius of the inner yoke (IY) and the inner radius of the outer yoke (OY), are considered equipotential, whose magnetomotive forces (MMFs) are set $F$ and 0. The radii of IY and OY are $r_1$ and $r_2$ respectively, and their centers are separated by a distance $d$. The coil (radius $r_c$) is located in the air gap between IY and OY.

The model of magnetic monopoles was introduced in 1931 by Dirac in his famous work, i.e., Dirac quantization condition \cite{dric}. It assumes the existence of a magnetic field density, yielding magnetic charges measured in SI unit A$\cdot$m. By introducing the magnetic charges, Maxwell's equations become beautifully symmetrical and a magnetic boundary problem can be equivalent to a conductive boundary problem in electrostatics \cite{new}. Therefore, the method of image can be employed here to solve the magnetic field in the air gap region.
Without losing generality, the $x$ axis is set along the maximum/minimum air gap, and the $y$ axis coincides with the electrical axis. Two point monopole charges, $+\rho$ and $-\rho$, are placed in mirror at $(b,0)$ and $(-b,0)$ where $b$ is an unknown variable. The distance of OA and OB, $h_1$ and $h_2$, are also unknown quantities. Based on the image method, we obtain \cite{electricalaxis}
\begin{equation}
r_1^2+b^2=h_1^2,~~~~r_2^2+b^2=h_2^2,~~~~h_2-h_1=d.
\label{eq.d}
\end{equation}
The unknown $b$ can be solved in equation (\ref{eq.d}), i.e.,
\begin{equation}
b=\sqrt{\left(\frac{r_2^2-r_1^2-d^2}{2d}\right)^2-r_1^2}.
\end{equation}

Ignoring all boundaries in figure \ref{fig2} and replacing the whole region with air, the magnetic potential $\varphi$ on the $xy$ plane can be calculated as
\begin{equation}
\varphi(x,y)=\frac{\rho}{4\pi\mu_0}\ln{\frac{(x+b)^2+y^2}{(x-b)^2+y^2}},
\end{equation}
where $\mu_0$ is the permeability of the vacuum (air). The magnetic field $H=B/\mu_0$ is the partial derivative of the magnetic potential $\varphi$, which can be calculated as
%\begin{equation}
%F=\frac{\rho}{4\pi\mu_0}\ln{\frac{(h_1+b)(h_2-b)}{(h_1-b)(h_2+b)}}
%\end{equation}
\begin{equation}
B_x=\mu_0\frac{\partial\varphi}{\partial x}=\frac{\rho b(y^2-x^2+b^2)}{\pi[(x-b)^2+y^2][(x+b)^2+y^2]},
\label{eq.bx}
\end{equation}
\begin{equation}
B_y=\mu_0\frac{\partial\varphi}{\partial y}=-\frac{2\rho bxy}{\pi[(x-b)^2+y^2][(x+b)^2+y^2]}.
\label{eq.by}
\end{equation}

\subsection{Numerical simulation of $Bl$}
As is known, the geometrical factor $Bl$ is defined as
\begin{equation}
Bl=\int_{L}({\bf B}\times \ud{\bf l})\cdot {\bf k},
\label{e17}
\end{equation}
where $L$ is the integral path along the coil, and ${\bf k}$ is a unit vector. In our analysis, the coordinate of the coil center, i.e., point C shown in figure \ref{fig2}, is set $(x_0, y_0)$. For an arbitrary point P on the coil, its coordinate can be expressed as $(x_0+r_c\cos\theta, y_0+r_c\sin\theta)$ where $\theta$ is the angle of PC$x$. According to equations (\ref{eq.bx}) and (\ref{eq.by}), two magnetic components at point P are calculated as
\begin{eqnarray}
B_x(P)=\frac{\rho b[(y_0+r_c\sin\theta)^2-(x_0+r_c\cos\theta)^2+b^2]}{\pi\{[(x_0+r_c\cos\theta)-b]^2+(y_0+r_c\sin\theta)^2\}}\nonumber\\
\times\frac{1}{[(x_0+r_c\cos\theta)+b]^2+(y_0+r_c\sin\theta)^2},
\label{eq.bxp}
\end{eqnarray}
\begin{eqnarray}
B_y(P)=-\frac{2\rho b(x_0+r_c\cos\theta)(y_0+r_c\sin\theta)}{\pi\{[(x_0+r_c\cos\theta)-b]^2+(y_0+r_c\sin\theta)^2\}}\nonumber\\
\times\frac{1}{[(x_0+r_c\cos\theta)+b]^2+(y_0+r_c\sin\theta)^2}.
\label{eq.byp}
\end{eqnarray}
Since the vector $\ud{\bf l}(P)=r_c(-\sin\theta, \cos\theta)\ud\theta$, $Bl$ can be rewritten as
\begin{equation}
Bl=\int_{0}^{2\pi}r_c[B_x(P)\cos\theta+B_y(P)\sin\theta]\ud\theta.
\end{equation}

\begin{figure}[tp!]
 \centering
 \includegraphics[width=0.45\textwidth]{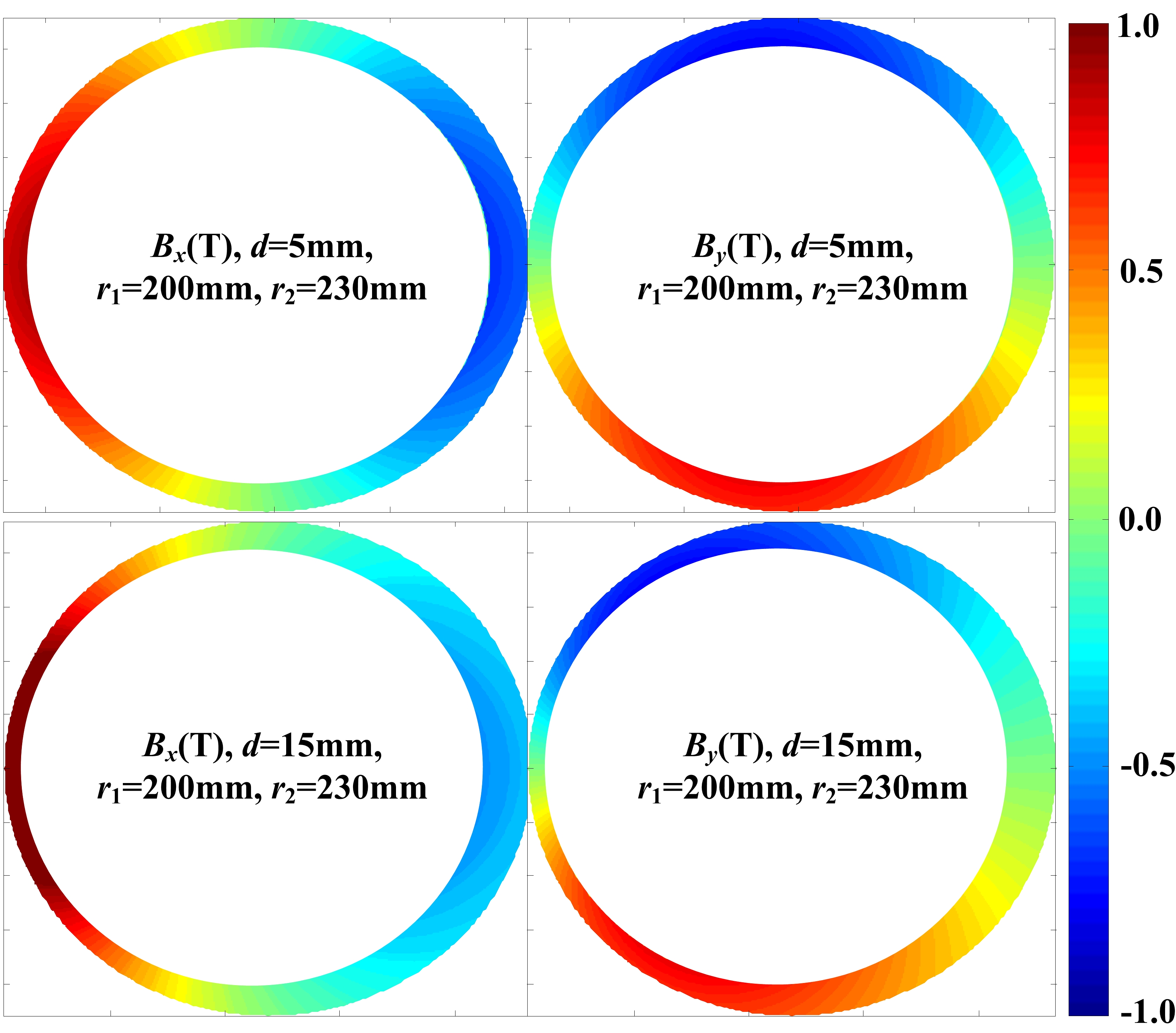}
 \caption{The magnetic flux density distribution of $B_x$ and $B_y$ in the air gap when $d=5$\,mm and $d=15$\,mm. In the calculation, $r_1$ and $r_2$ are set as 200\,mm and 230\,mm.}
   \label{fig3}
\end{figure}

To be sample, numerical simulation of the magnetic field distribution and $Bl$ calculation is demonstrated in several examples. Figure \ref{fig3} shows the numerical simulation of the magnetic flux density distribution of $B_x$ and $B_y$ in the air gap when $d=5$\,mm and $d=15$\,mm. It is apparent that the magnetic field strength is inversely proportional to the air gap distance.

\begin{figure}[tp!]
 \centering
 \includegraphics[width=0.45\textwidth]{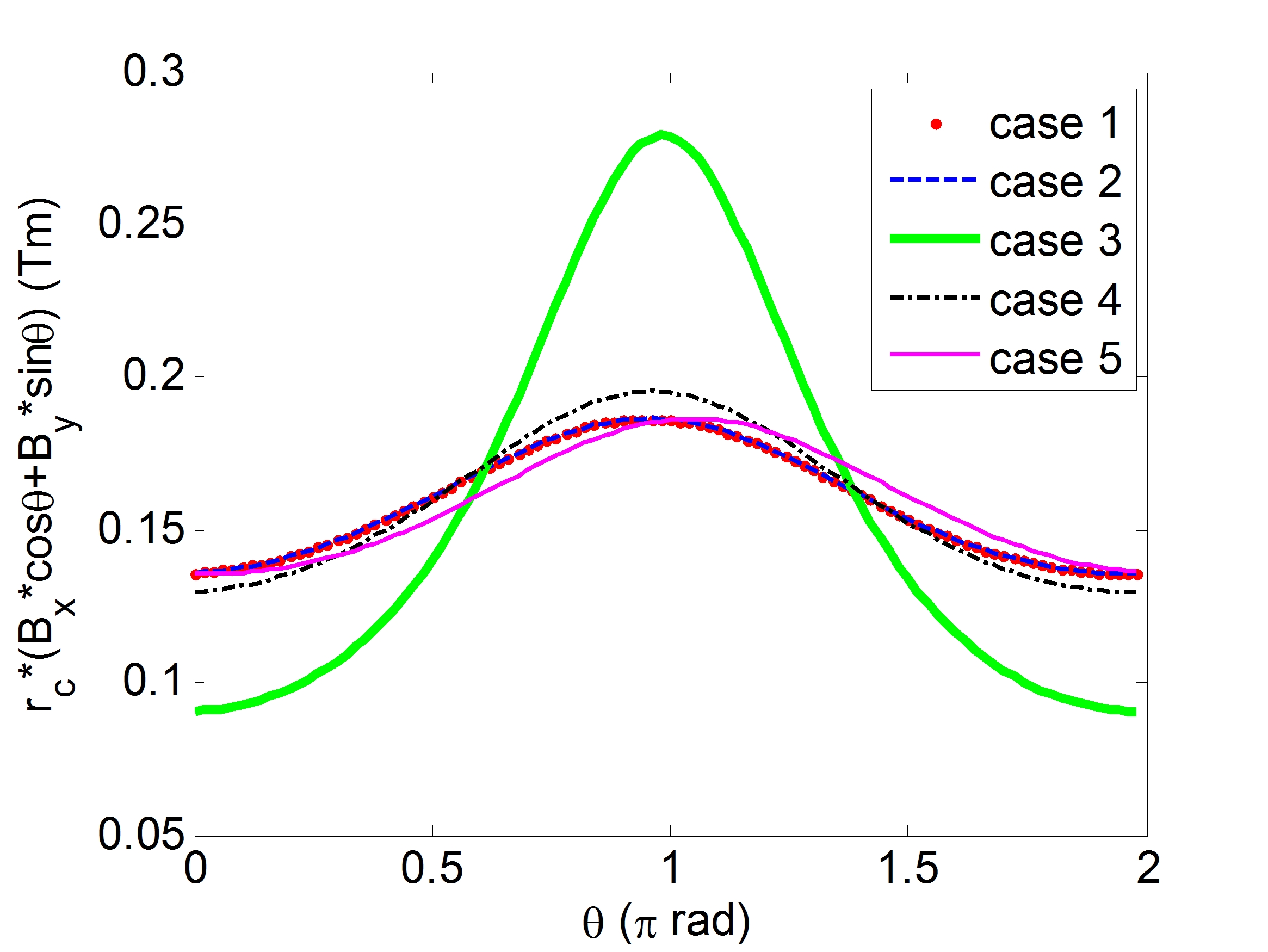}
 \caption{The calculation result of the integral function $\Delta$ as a function of $\theta$ in different parameter setups. The case number is the same as the number shown in table \ref{table1}.}
   \label{fig4}
\end{figure}

We select five cases with different parameters to show how the integral function, i.e., $\Delta=r_c[B_x\cos\theta+B_y\sin\theta]$, varies as a function of $\theta$. The parameters and $Bl$ calculation results are listed in table \ref{table1}, and $\Delta(\theta)$ are shown in figure \ref{fig4}. It is shown that compared to the ideal $1/r$ field case ($\Delta$ is a constant), the variation of $\Delta$ at different $\theta$ locations is completely neutralized, and the $Bl$ is shown in these examples to be constant and independent from a symmetrically changing the coil radius $r_c$, the IY and OY misalignment $d$ and the coil position $(x_0, y_0)$. Note that in the $Bl$ calculation, we use 500 steps for the numerical integration, and it is found that the result unvaried when smaller or larger integration steps are employed.
\begin{table}
\renewcommand{\arraystretch}{1.2}
% if using array.sty, it might be a good idea to tweak the value of
% \extrarowheight as needed to properly center the text within the cells
\caption{Parameters setup and calculation result of the simulation examples.}
\label{table1}
\centering
\begin{tabular}{cccccc}
\hline
Case No.&$r_c$\,(mm) & $d$\,(mm) & $x_0-h_1-d$\,(mm) & $y_0$\,(mm) & $Bl$\,(Tm)\\
\hline
1&215 & 5 & -5 & -5&1.00000000\\
2&220 & 5 & -5 & -5&1.00000000\\
3&215 & 15 & -5 & -5&1.00000000\\
4&215 & 5 & 5 & -5&1.00000000\\
5&215 & 5 & -5 & 5&1.00000000\\
\hline
\end{tabular}
\end{table}

\section{Extension and explanation of the $Bl$ conservation}
\label{sec4}
\subsection{Gauss's law on magnetism}
As a matter of fact, the $Bl$ conservation can be extended to any two dimensional magnetic field planes. Here we give a brief proof using the Gauss's law for magnetism with a model of magnetic monopoles.
Revising Gauss's law for magnetism \cite{new}, we have
\begin{equation}
\oint_{S}{\bf B}\cdot d{\bf A}=\frac{\rho_m}{\mu_0},
\label{e21}
\end{equation}
where $S$ denotes a closed surface, $d{\bf A}$ a vector whose magnitude is the area of an infinitesimal piece of the surface $S$, pointing outward, $\rho_m$ the theoretical magnetic charge and $\mu_0$ the permeability of free space.

\begin{figure}[tp!]
 \centering
 \includegraphics[width=0.3\textwidth]{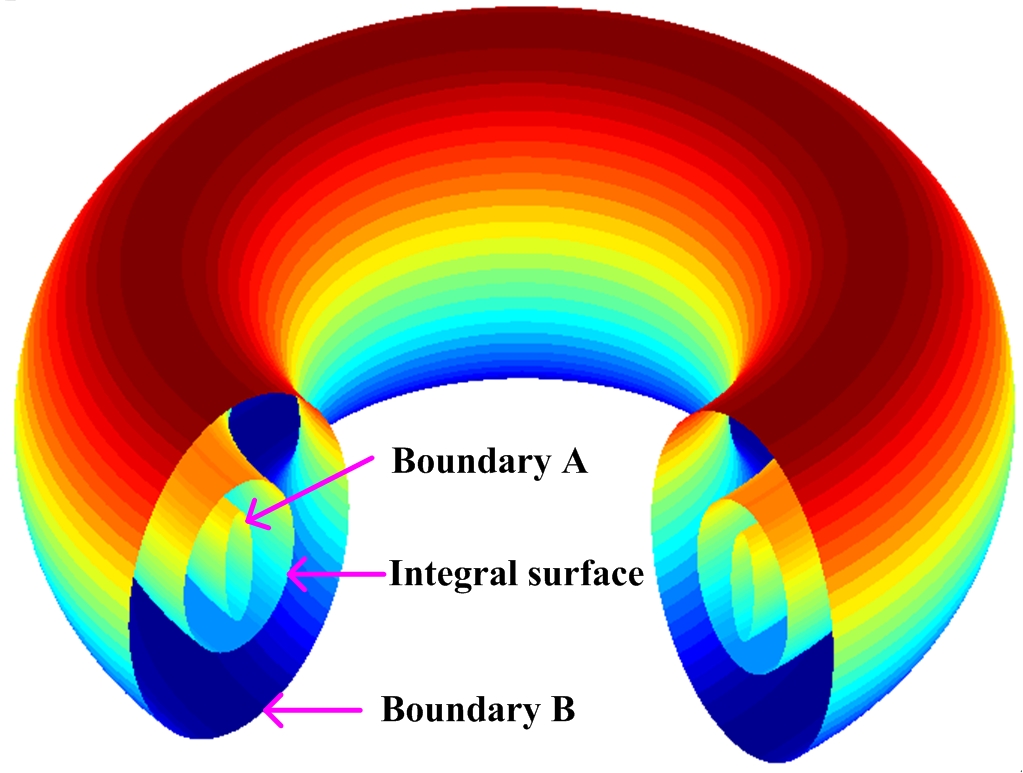}
 \caption{The three dimensional space obtained by rotating the two dimensional field plane (3/4 view).}
   \label{fig5}
\end{figure}

\begin{figure}[tp!]
 \centering
 \includegraphics[width=0.45\textwidth]{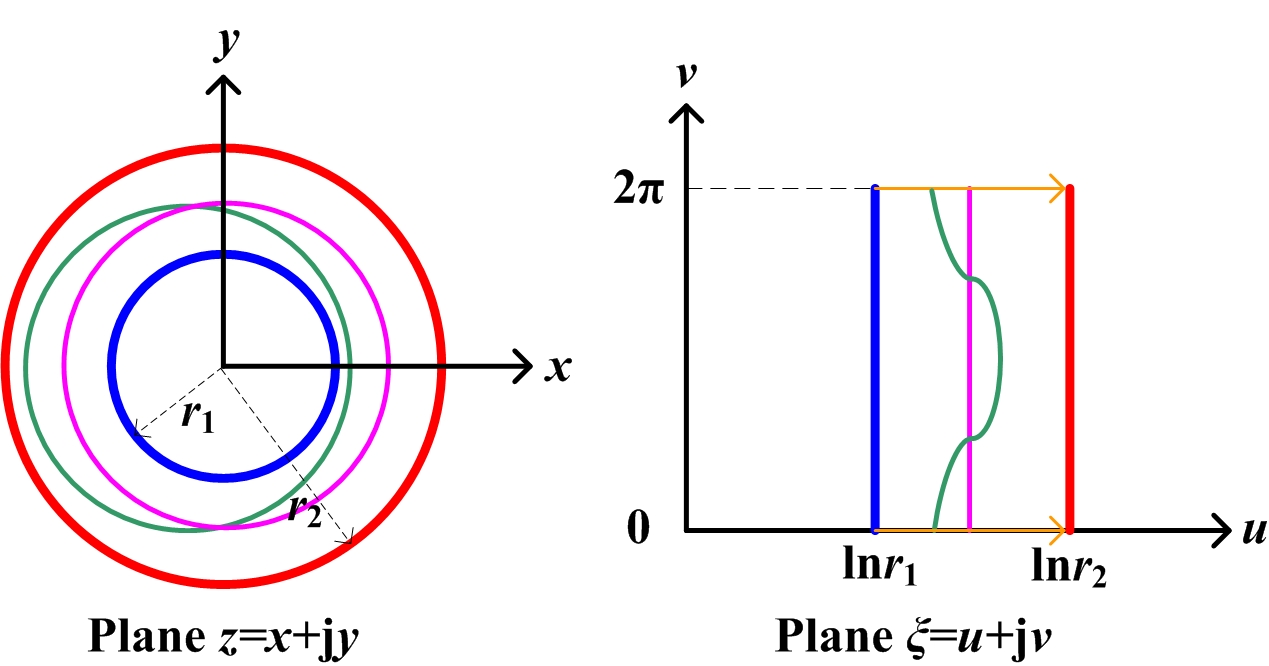}\\
 (a)~\\
  \includegraphics[width=0.45\textwidth]{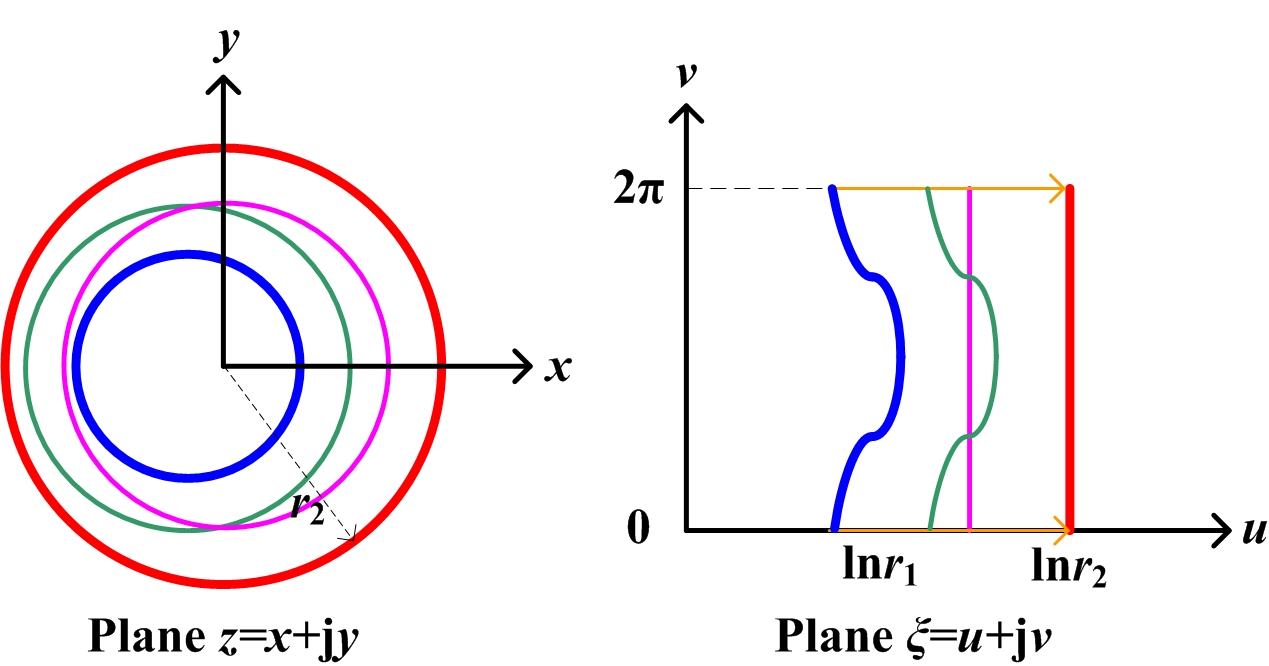}\\
  (b)~\\
 \caption{The conformal transformations of different alignment cases for concentrically aligning the IY and OY. (a) shows an ideal symmetrical case with different coil positions and (b) presents the misalignment case with a same setup of coil position. The blue line and red line denote IY and OY respectively. The pink line is the ideal coil position while the green line is an arbitrary coil position. On the $xy$ plane, the origin of coordinate is placed at the OY.}
   \label{fig6}
\end{figure}

It has been noticed that the $Bl$ is calculated in terms of a cross product in equation (\ref{e17}) rather than a dot product in equation (\ref{e21}). In order to apply the Gauss's law of magnetism, we firstly show how the $Bl$ is related to the magnetic flux in two dimensional vectors. ${\bf B}\times d{\bf l}$ can be expanded
\begin{eqnarray}
{\bf B}\times d{\bf l}=(B_x, B_y)\times(-r_c\sin\theta, r_c\cos\theta)\nonumber\\
=B_xr_c\cos\theta+B_yr_c\sin\theta,
\label{eq.flux}
\end{eqnarray}
while the dot product of the vector ${\bf B}$ and the vector $d{\bf n}=(r_c\cos\theta, r_c\sin\theta)$ ($d{\bf n}$ is the normal of $d{\bf l}$) is calculated as
\begin{eqnarray}
{\bf B}\cdot d{\bf n}=(B_x, B_y)\cdot(r_c\cos\theta, r_c\sin\theta)\nonumber\\
=B_xr_c\cos\theta+B_yr_c\sin\theta.
\label{eq.dot}
\end{eqnarray}
A comparison of equations (\ref{eq.flux}) and (\ref{eq.dot}) yields ${\bf B}\times d{\bf l}={\bf B}\cdot d{\bf n}$, which shows in a two dimensional field plane $Bl$ can be calculated by integrating the magnetic flux along the coil.

Secondly, we rotate the two dimensional magnetic field plane along a uniform circle, yielding a three dimensional space shown in figure \ref{fig5}. In the watt balance case, boundaries A and B are formed by rotating lines IY and OY while the integral surface is created by the coil. Figure \ref{fig5} is a typical expression of Gauss's law.
Assuming the total magnetic charge on the boundary A is $\rho_m$, the geometrical factor $Bl$ is determined as
\begin{equation}
Bl=\frac{\rho_m}{2\pi\mu_0}.
\label{eq.fine}
\end{equation}

Equation (\ref{eq.fine}) shows the $Bl$ of a watt balance coil is only determined by the magnetic charge $\rho_m$, independent of the yoke shape, the alignment of the inner and outer yokes, the coil position and the coil shape.

\subsection{Conformal transformation}
 The conformal transformation, or the conformal mapping, has an excellent property to modify only the geometry of a polygonal structure, preserving its physical magnitudes \cite{ct}. The $Bl$ conservation in a two dimensional magnetic field plane can also be simply explained by conformal transformations. The following conformal transformation is applied, i.e.,
\begin{equation}
\xi=\ln z,
\label{eq.ct}
\end{equation}
where $z=x+\mbox{j}y$ denotes the $xy$ plane, and $\xi=u+\mbox{j}v$ is the transformation plane. Figure \ref{fig6} presents the conformal transformations of the ideal symmetrical and the asymmetrical cases for concentrically aligning the IY and OY. Note that there is no fringe effect in the conformal transformation of equation (\ref{eq.ct}). It can be seen that the transformation on the $\xi$ plane is comparable to a capacitor with parallel or non parallel plates. Since the $Bl$ is proportional to flux through the coil in the normal direction, for a two plate capacitor, the flux through the coil (the line between two plate lines) should be a constant only related to the charge of the high potential plate. Therefore, $Bl$ is independent to the yoke misalignments in $r$ direction and the coil horizontal displacements.

\section{Conclusion}
\label{sec5}
For watt balances, the $Bl$ is independent from horizontal coil displacements in a $1/r$ magnetic field but the $1/r$ symmetry of the magnetic field strongly depends on the alignment of the IY and OY. We presented an analytical calculation of the magnetic field distribution when the $r$ symmetry of the magnet is broken, i.e., IY and OY are not concentrically aligned. The analysis shows that $Bl$ conservation is independent of the $r$ symmetry of the magnetic field in the neutral plane $z=0$. The $Bl$ conservation established on any arbitrary, pure two dimensional magnetic plane, is proven by Gauss's law of magnetism or conformal transformations. This generalized $Bl$ conservation analysis can provide insight for a much more relaxed construction and alignment procedure for watt balance magnets.

\end{document}